\documentstyle[epsfig,12pt,english,here,oldlfont]{article}
\begin{document}
\baselineskip=0.8cm
\newcommand{\ini}{\begin{equation}}
\newcommand{\fin}{\end{equation}}
\newcommand{\inir}{\begin{eqnarray}}
\newcommand{\finr}{\end{eqnarray}}
\newcommand{\inif}{\begin{figure}}
\newcommand{\finf}{\end{figure}}
\newcommand{\bc}{\begin{center}}
\newcommand{\ec}{\end{center}}
\def\ol{\overline}
\def\pa{\partial}
\def\ra{\rightarrow}
\def\ts{\times}
\def\df{\dotfill}
\def\bs{\backslash}

$~$

\hfill DSF-T-98/21

\vspace{1 cm}

\centerline{\LARGE{Relation between quark masses}}

\centerline{\LARGE{and weak mixings}}

\vspace{1 cm}

\centerline{\large{D. Falcone$^{*+}$ and F. Tramontano$^*$}}

\vspace{1 cm}

\centerline{$^*$Dipartimento di Scienze Fisiche, Universit\`a di Napoli,}
\centerline{Mostra d'Oltremare, Pad. 19, I-80125, Napoli, Italy;}
\centerline{$^+$INFN, Sezione di Napoli, Napoli, Italy}

\centerline{e-mail: falcone@na.infn.it}
\centerline{e-mail: tramontano@na.infn.it}

\vspace{1 cm}

\centerline{ABSTRACT}

Simple transformation formulas between fermion matrices and observables, and
numerical values of quark matrices, are obtained on a particular weak basis
with one quark matrix diagonal and the other with vanishing elements
1-1, 1-3 and 3-1, and with only the element 2-2 complex.
When we choose $M_u$ diagonal, then $M_d$ shows intriguing numerical
properties which suggest a four parameter description of it,
which implies $V_{us}\simeq \sqrt{m_d/m_s}$,
$V_{cb}\simeq (3/\sqrt{5})(m_s/m_b)$ and
$V_{ub}\simeq (1/\sqrt{5})(\sqrt{m_d m_s}/m_b)$.
Few comments on mass-mixing relations are added.

PACS numbers: 12.15.Ff, 12.15.Hh

\newpage

In the standard model Lagrangian \cite{lan},
written in a general weak basis, quark mass matrix elements are not
explicitly related to physical observables, that is quark masses and
weak mixings.
The problem of finding such a relation, without extra symmetries, has been
addressed in \cite{blm,ho,kt,fpr}.
In particular, in \cite{blm} it was shown that it is always possible to find
a weak basis where the quark mass matrices have the nearest neighbor
interaction form and depend on twelve real parameters.
Two of these twelve parameters are arbitrary \cite{ho} and
related to the phase convention of the weak
mixing matrix \cite{kt}. Then, in \cite{fpr},
it was shown that it is always possible to set one quark matrix in the
diagonal form and the other in a form with zero entries in positions
1-1, 2-2 and 3-1, and with only the element 1-2 complex.
In such a way mass matrices contain ten real parameters, exactly the same
number of physical observables, six quark masses and three mixing angles
and one phase. This corresponds to the choice of a
$minimal$ $parameter$ $basis$ \cite{k1}. As one mass matrix is chosen to be
diagonal, it is relatively easy to obtain exact transformation formulas
between mass matrices and observables. Other minimal parameter bases are
considered in \cite{fk,hs}. Here we describe a further minimal parameter basis,
which shows interesting properties and on which transformation formulas are
simple.

In fact it is also always possible \cite{marosa} to choose a weak basis for which
\ini
M_d=diag(m_d,m_s,m_b)
\fin
and
\ini
M_u=\left( \begin{array}{ccc}
            0 & M_{12} & 0 \\
            M_{21} & M_{22} & M_{23} \\
            0 & M_{32} & M_{33}
            \end{array} \right)
\fin
(or $M_u$ is diagonal and $M_d$ has the form (2)).

On this basis the relation between mass matrices and observables is given by
\ini
M_u M_u^+=K^+ \cdot diag(m_u^2,m_c^2,m_t^2) \cdot K \equiv X^u
\fin
where $K$ is the Cabibbo-Kobayashi-Maskawa (CKM) matrix \cite{ckm}.
In the case of $M_u$ diagonal we have instead
\ini
M_d M_d^+=K \cdot diag(m_d^2,m_s^2,m_b^2) \cdot K^+ \equiv X^d.
\fin
By writing $M_{ij}=m_{ij}e^{i r_{ij}}$ we can reconstruct the usual
representations of $K$ \cite{rpp}
by means of three non vanishing phases $r_{12}$, $r_{22}$ and $r_{23}$.
The product $M_u M_u^+$ is then given by
\ini
\left( \begin{array}{ccc}
            m_{12}^2 & m_{12}m_{22}e^{i(r_{12}-r_{22})} &
            m_{12}m_{32}e^{ir_{12}} \\
            m_{12}m_{22}e^{-i(r_{12}-r_{22})} & m_{21}^2+m_{22}^2+m_{23}^2 &
            m_{22}m_{32}e^{ir_{22}}+m_{23}m_{33}e^{ir_{23}} \\
            m_{12}m_{32}e^{-ir_{12}} &
            m_{22}m_{32}e^{-ir_{22}}+m_{23}m_{33}e^{-ir_{23}} &
            m_{32}^2+m_{33}^2
            \end{array} \right).
\fin
and the trasformation formulas between masses and mixings in $X$ and mass
matrix elements in $M$ are written in a very simple form
\ini
m_{12}=\sqrt{X^u_{11}}
\fin
\ini
m_{22}=|X^u_{12}|/m_{12}
\fin
\ini
m_{32}=|X^u_{13}|/m_{12}
\fin
\ini
m_{33}=\sqrt{X^u_{33}-m_{32}^2}
\fin
\ini
r_{12}=phase(X^u_{13})
\fin
\ini
r_{22}=r_{12}-phase(X^u_{12})
\fin
$$
M_{23}=(X^u_{23}-m_{22}m_{32}e^{ir_{22}})/m_{33}
$$
\ini
m_{23}=|M_{23}|
\fin
\ini
r_{23}=phase(M_{23})
\fin
\ini
m_{21}=\sqrt{X^u_{22}-m_{22}^2-m_{23}^2}.
\fin
In the case of $M_u$ diagonal the same formulas hold with $X^u \ra X^d$.
With a phase transformation of quark fields,
\ini
M_{u,d} \ra diag(e^{-ir_{12}},e^{-ir_{23}},1) \cdot M_{u,d} \cdot
diag(e^{ir_{23}},1,1), 
\fin
only a phase $r'_{22}=r_{22}-r_{23}$ remains in the element 2-2,
and we obtain, using numerical values of quark
masses at $\mu=M_Z$ as in \cite{fk} ($m_u=0.00233$, $m_c=0.677$,$m_t=181$,
$m_d=0.00469$, $m_s=0.0934$, $m_b=3.00$ $GeV$) and mixings
as in \cite{rpp} (with $\delta=1.35$),
\ini
M_u=\left( \begin{array}{ccc}
            0 & 1.591 & 0 \\
            0.011 & 7.118~e^{1.334i} & 0.269 \\
            0 & 180.1 & 17.02
            \end{array} \right)~GeV,
\fin
and if instead we choose $M_u$ to be diagonal,
\ini
M_d=\left( \begin{array}{ccc}
            0 & 0.024 & 0 \\
            0.021 & 0.105~e^{-1.205i} & 0.106 \\
            0 & 1.333 & 2.685
            \end{array} \right)~GeV.
\fin
We can see that in (16), due to the large value of the top quark mass,
the biggest matrix element is not 3-3, as in (17),
but the element 3-2. This feature is different from the basis in \cite{fpr}
where the biggest element is the element 3-3 either if $M_u$ or $M_d$ is 
diagonal. Moreover, the numerical values in (17) suggest to take
$M_u$ diagonal and
\ini
M_d=\left( \begin{array}{ccc}
            0 & a & 0 \\
            a & b~e^{i\varphi} & b \\
            0 & c & 2c
            \end{array} \right),
\fin
where $a$, $b$ and $c$ are of order $10^{-2}$, $10^{-1}$ and $1$ $GeV$,
respectively.
From (18) we obtain the approximate expression
\ini
M_d\simeq \left( \begin{array}{ccc}
            0 & \sqrt{m_d m_s} & 0 \\
            \sqrt{m_d m_s} & m_s~e^{i\varphi} & m_s \\
            0 & m_b/\sqrt{5} & 2m_b/\sqrt{5}
            \end{array} \right).
\fin
In the heavy quark limit $m_b \gg m_s,m_d$ we have the effective matrix
for the two lightest down quarks
\ini
M_d \simeq \left( \begin{array}{cc}
            0 & \sqrt{m_d m_s} \\
            \sqrt{m_d m_s} & m_s
            \end{array} \right)
\fin
which gives the famous relation \cite{gstcm,wein}
\ini
V_{us}\simeq \sqrt{\frac{m_d}{m_s}}.
\fin
In the chiral limit $m_d \ll m_s,m_b$ we have instead, for the two heaviest
down quarks
\ini
M_d \simeq \left( \begin{array}{cc}
            m_s & m_s \\
            m_b/\sqrt{5} & 2m_b/\sqrt{5}
            \end{array} \right)
\fin
and when we diagonalize the Hermitian matrix
\ini
M_d M_d^+ \simeq \left( \begin{array}{cc}
            m_s^2 & 3m_s m_b/\sqrt{5} \\
            3m_s m_b/\sqrt{5} & m_b^2
            \end{array} \right)
\fin
we obtain the relation
\ini
V_{cb}\simeq \frac{3}{\sqrt{5}}\frac{m_s}{m_b},
\fin
which gives $V_{cb}=0.042$ to be compared with the experimental value
$0.041 \pm 0.005$ \cite{rpp}.
Finally, taking the full matrix
\ini
M_d M_d^+ \simeq \left( \begin{array}{ccc}
            m_d m_s & m_s \sqrt{m_d m_s}e^{-i\varphi} & m_b\sqrt{m_d m_s/5} \\
            m_s \sqrt{m_d m_s}e^{i\varphi} & m_s(m_d+2m_s) &
             {m_s m_b}(e^{i\varphi}+2)/\sqrt{5} \\
            m_b\sqrt{m_d m_s/5} &
            m_s m_b(e^{-i\varphi}+2)/\sqrt{5} & m_b^2
            \end{array} \right)
\fin
we have the relation
\ini
V_{ub}\simeq \frac{1}{\sqrt{5}}\frac{\sqrt{m_d m_s}}{m_b},
\fin
which gives $V_{ub}=0.003$ to be compared with the experimental
range $0.002\div 0.005$ \cite{rpp}.
From (21),(24) and (26) we yield also
\ini
\frac{V_{ub}}{V_{cb}}\simeq \frac{1}{3}V_{us}.
\fin

Setting $x=\sqrt{m_d/m_s}$ and $y=\sqrt{m_s/m_b}$, we have
$V_{us}\simeq x$, $V_{cb}\simeq (3/\sqrt{5})y^2$ and
$V_{ub}\simeq (1/\sqrt{5})xy^2$ which means that, on this basis, weak mixings,
apart from numerical coefficients not so different from 1,
are generated by square roots of
quark mass ratios $x$ and $y$. Of course $x\sim y\sim \lambda$ leads to
the Wolfenstein parametrization \cite{wol} of the
CKM matrix. On the basis with $M_d$ diagonal and $M_u$ given by (2)
such simple features are lost. Nevertheless weak mixings appear related
to up quark ratios (for example $V_{cb}\simeq 11~m_c/m_t$). Then, from
the paper \cite{hw}, where the relations $V_{us} \simeq \sqrt{m_d/m_s}$,
$V_{cb} \simeq m_s/m_b$, but $V_{ub}\simeq \sqrt{m_u/m_t}$ were
inferred, and our work, we argue that choosing different weak bases
we can accordingly obtain different relations between mixings and masses,
each of them in agreement with experimental data. Hence, each weak basis
may be useful to describe some features of fermion masses and mixings.
As a last remark on the basis considered here, we observe that, as written
in footnote 6 of \cite{blm}, in left-right symmetric models \cite{ps}
both $M_u$ and $M_d$ can always take the form (2).

In conclusion, we have obtained very simple formulas for relating fermion 
matrices to observables, and numerical values of quark matrices on a basis
with one quark mass matrix diagonal and the other with three zeros in positions
1-1, 1-3 and 3-1. Such numerical values suggest a simple
form for $M_d$ which does imply relations (21),(24),(26), and (27).
Moreover, on this basis weak mixings have a simple expression. 

$~$

We thank F. Buccella for useful comments and L. Rosa and O. Pisanti
for suggestions and many discussions.

\end{document}